# Simulating the gravity in $S_{11}$ parameters and friis transmission equation

Sunit Shantanu Digamber Fulari, Department of electronics and communication, Chandigarh University, Assistant professor, Govt college of Arts, Science and Commerce, Khandola, Marcela, Goa

**Abstract:** Antennas are taking design shapes by the orientation of its material and the final structural design they take. Irrespective of the shape, the function and efficiency they produce does not change and vary to much extent. We are trying to simulate a design of antenna which is using the artificial intelligence model to Specific absorption rate minimization. We have successfully used antenna magus and CST suite to produce our improved simulation results. Motion models such as optical flow is used to smartly detect communicating devices inorder to connect and link the devices. The algorithm used by us is closest neighbour algorithm which is connected with antenna to apply it in object detection of mobile devices and antennas. We are trying to have a six second break in which the antenna will operate at minimum frequency but at the same time operating in its functioning. After every six seconds the antenna will radiate its frequency of vibration to detect if there is any new object or mobile device in sight.

**1. Introduction:** We are trying to simulate a differently shaped antenna design which will eliminate noise and maximize the power transmitted by the antenna. We want the antenna to function causing minimum radiation in the defined direction, it will function with a old closest neighbour algorithm. We have also tried to use travelling salesman problem in order to simulate our defined working of antenna system. We have tried to make the antenna function in every six seconds at maximum power while the rest of the time it will function will least power required for it to function at that frequency in order to sustain the working of the antenna system.

**2. Material used for experimentation:** We have used a copper annealed with pure copper outer ring as a material. This is a very distinct design of the material which will produce some effects in every six seconds of its operation. We will try to define a circular design in circular form which will radiate energy in all directions though the shape of the antenna is immaterial to our functioning.

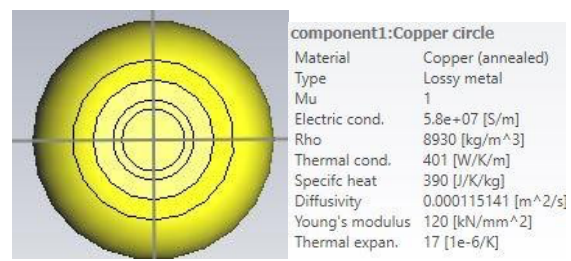

Figure 1:

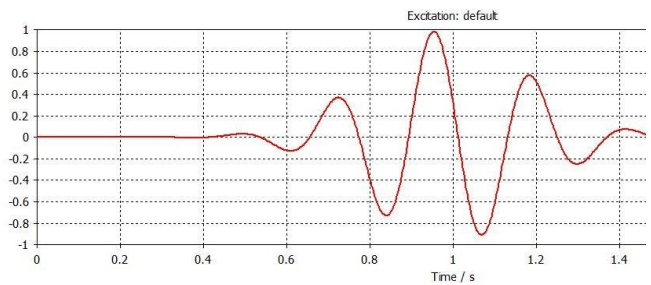

Figure 2: Excitation

Excitation signals of the given given antenna material is given by the above graph.

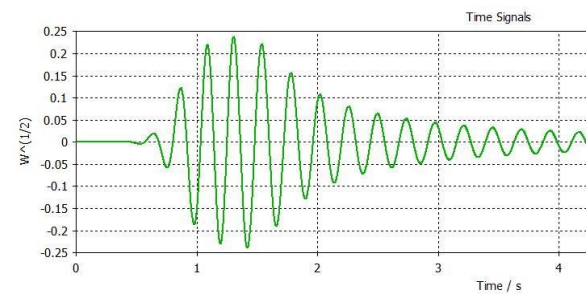

Figure 3:Portability signal is given by the above graph. These graphs are very clear in our simplified design which justify the functioning of the waves in order to produce ripple kind of effect. The frequency does fluctuate very distinctly but then later on becomes to nullifiable value. We are trying to simulate these copper annealed malleable wires and structure in order to produce our antenna synthesis. Furthermore we have developed various other simulations which will be covered in the later part of the paper.

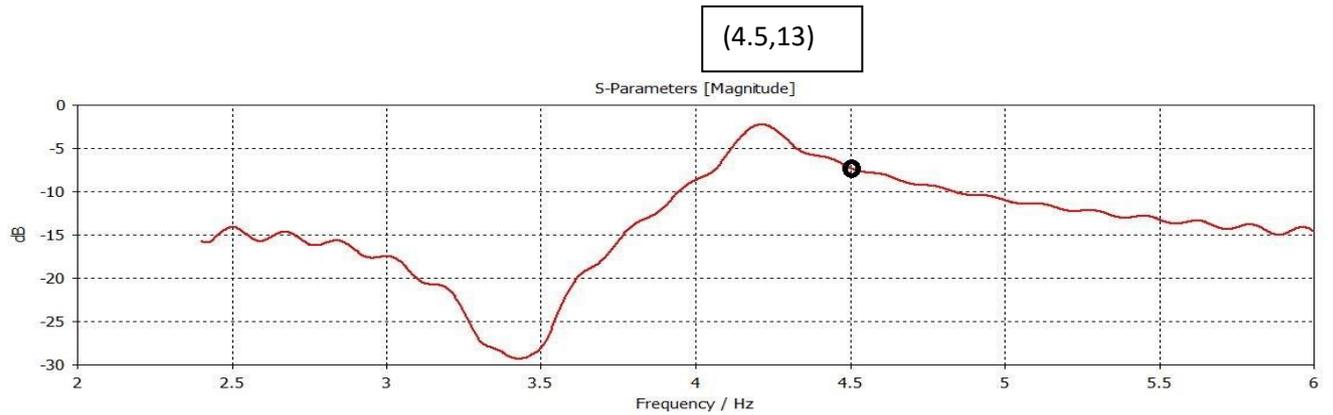

Figure 4:Scattering diagram for various electrical impulses given as above diagram.

S parameters are given as above.

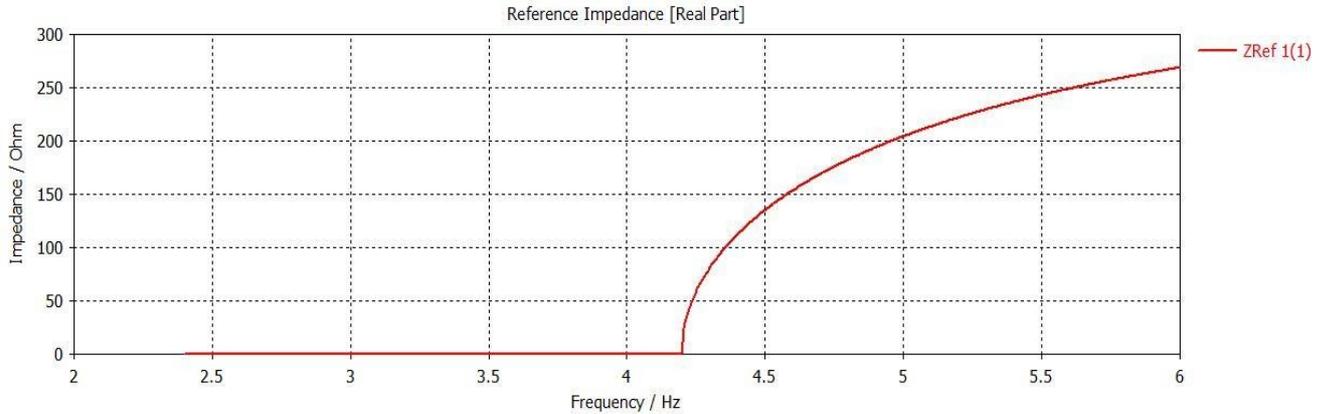

Figure 5:Reference impedance of s parameter is given as under.

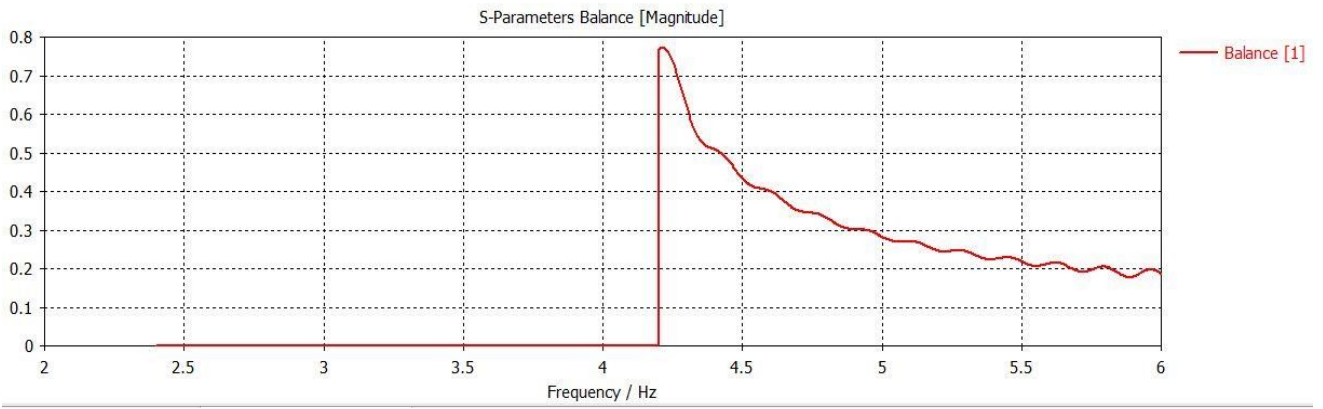

Figure 6:Balance in our design is given as above.

These simulations define some of the functioning of our antenna design which is highly malleable though its immaterial but has quite a proper significance. These parameters together define a very subjective antenna synthesis which is very important in our observation of study.

Secondly We have also used FR4 lossy and PEC as the next material similar to design used in copper annealing.

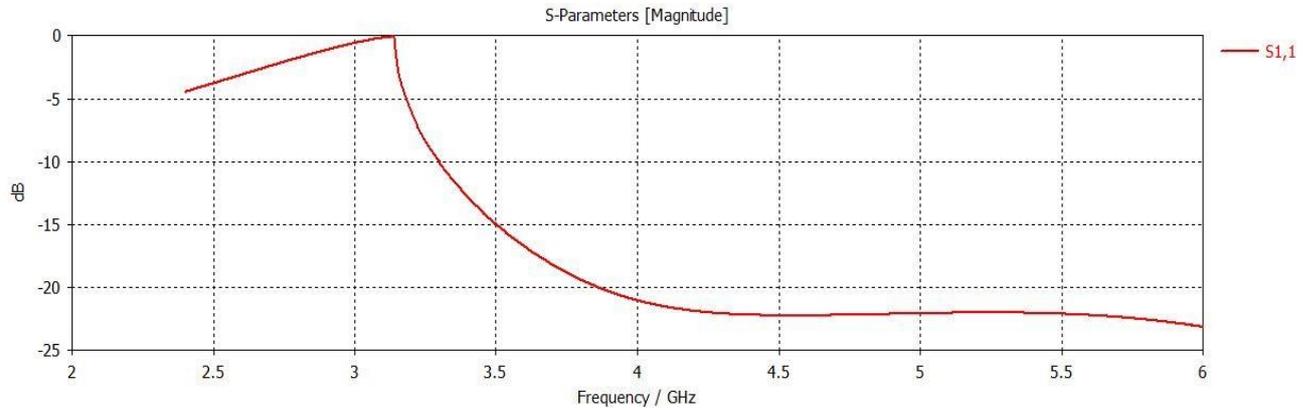

Figure 7:shows the s parameter of the design for increasing frequency.

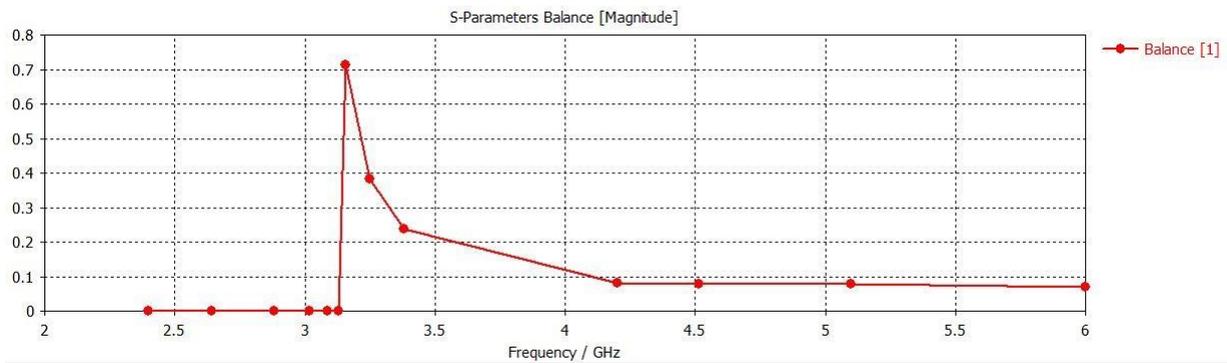

Figure 8: Giving the balance of the s parameters abtained as before.

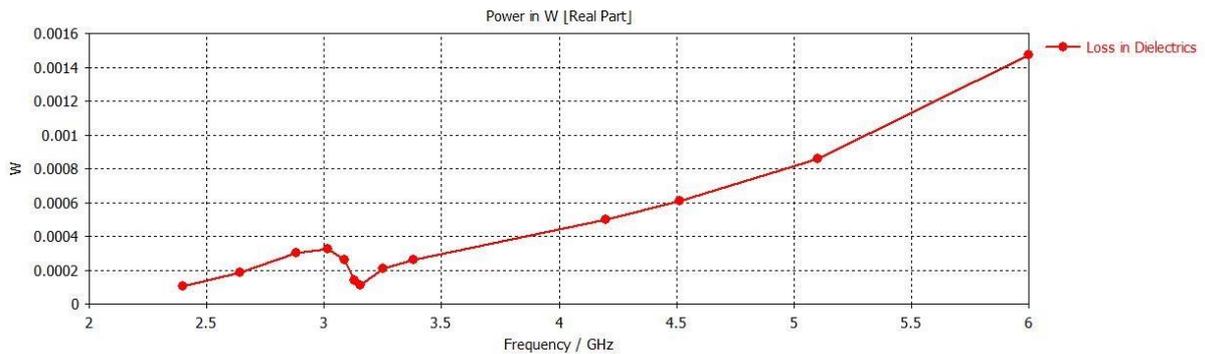

Figure 9:Loss in dielectrics is given as below. This is the loss in energy during electrification of the material.

So the figure 8 and 9 gives us a steady balance and loss in dielectric which increases with increase in frequency.

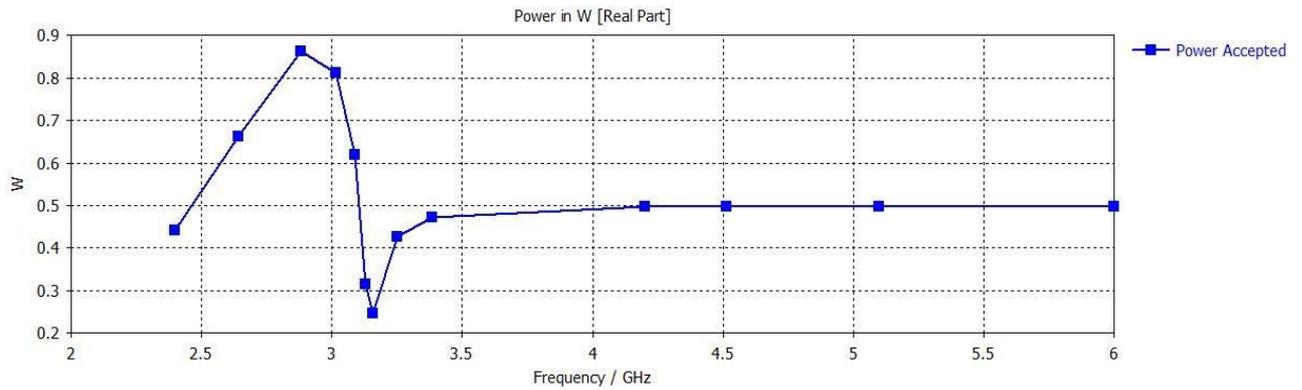

Figure 10 giving the power accepted in the material is power stimulated minus power outgoing.

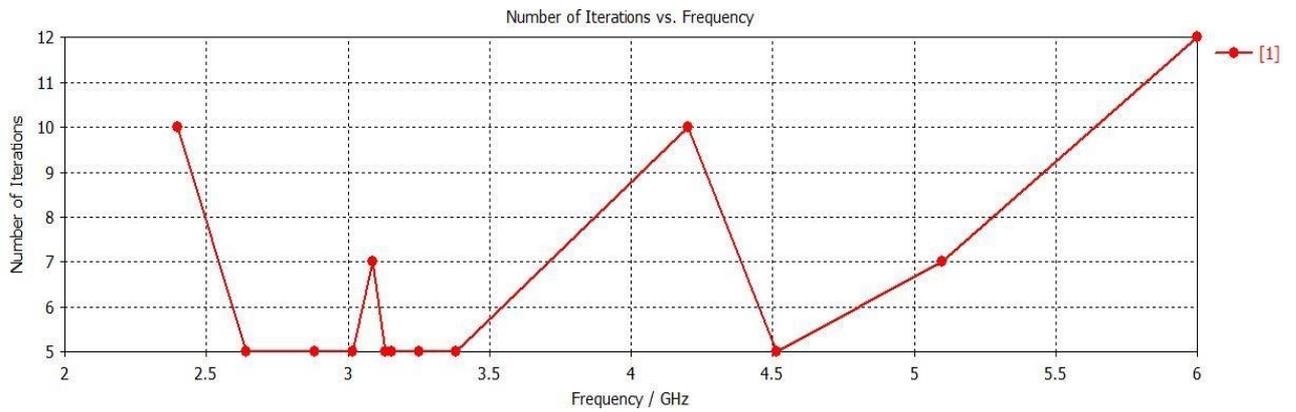

Figure 11: Iterations in the figure is given as above where the variations are predominantly defined.

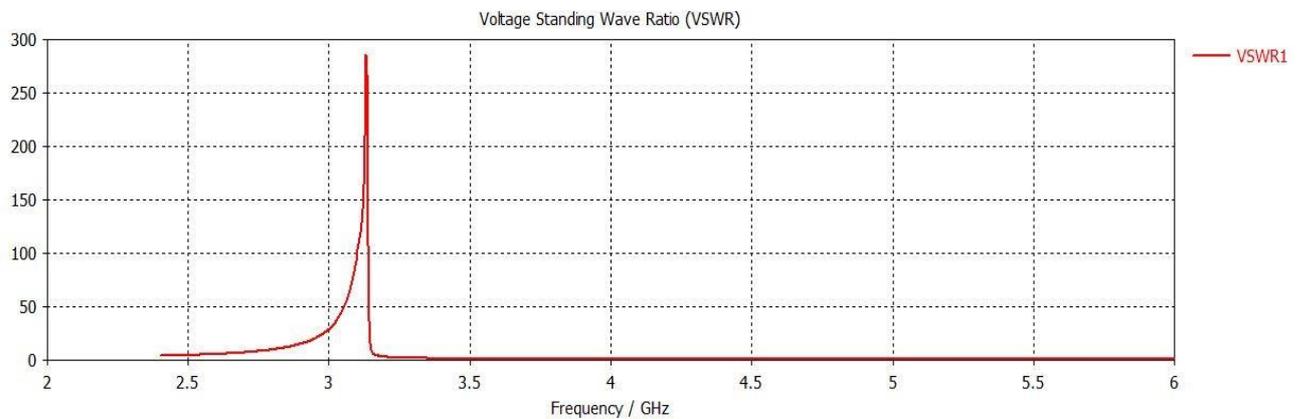

Figure 12 is the transmitted to the reflected wave energy in the wave. Which effectively shows the advantage of the material in its latency of operation. This device will be used in an interval of 6 seconds time break will degenerate the radiation to the minimum energy. This will inreturn produce our hypothesis and phenomenon of nearest neighbour algorithm in antenna array synthesis. Our proposed method involves 5G towers which have more than 15-20 kilometres of intersecting connectivity between towers and devices. Each tower in turn connects with each other by a tower of antenna which help in mobile communication. We have proposed a city connectivity transacting connection of antennas between 'margao', 'ponda' and 'panaji' connecting for 5G mobile communication using k nearest neighbour algorithm. There are spacing connecting between the towers shown in the below figure.

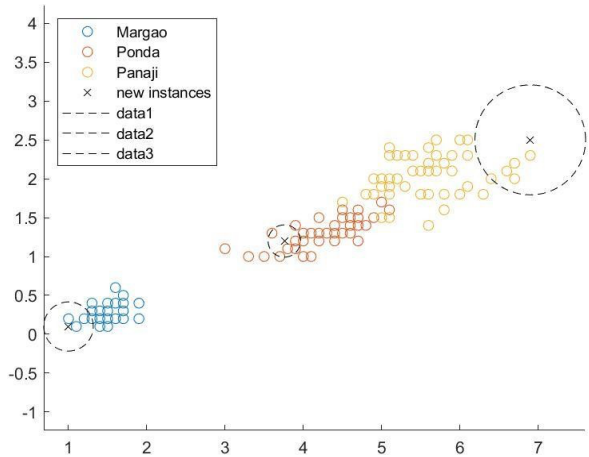

Figure 13: The above is intersecting towers connecting between three cities in state of Goa.

We will study the Friis transmission equation for clarity in our design of wave equations. We know the field is given by kq/r. The distance is very great that is R>>2$D^2_{max}$/λ. Which is in our case separated by 30-40 kilometres apart. Power density is given by

$$w = \frac{P_t}{4\Pi R^2} G_t(\theta_t, \varphi_t)$$

$$w = \frac{P_t}{4\Pi R^2} e_t D_t(\theta_t, \varphi_t)$$

$e_t$ and $D_t$ being the radiation efficiency and directivity of the antenna system.

$P_r$ is the power at the receiving antenna given by AW. as below

$P_r = A_r W_t$

Radiation efficiency and polarization loss factor are added together.

$P_r = e_r \cdot PLF \cdot A_r W_t = A_r W_t e_r |\hat{\rho}_t \cdot \hat{\rho}|^2$

Impedance and polarization matched equation for transmitting and receiving antenna is given by.

$$\frac{P_r}{P_t} = \left(\frac{\lambda}{4\Pi R}\right)^2 G_t(\theta_t, \varphi_t) G(\theta_r, \varphi_r)$$

The above equation is the famous friis transmission equation. Free-space loss factor being in first bracket.

Received power $P_r$

We obtain the transmitted power and gain of the antenna multiplied together.

$P_r = P_t G_t e_{TL}$

This term being the loss efficiency of the transmission line . $e_{TL}$.

Power radiated is given by the above equation as the maximum power radiated by the antenna given in terms of isotropic radiator would emit when producing the peak power density in the observed direction of maximum power radiation

$$P_r = 4\Pi U_{max}.$$

These friis transmission equation and power radiated play a significantly important role in our research paper. Power would be radiated every six seconds in

3D dimension in all directions. So the antennas would operate in the given interval of 6 seconds and even operate in full functioning but there will not be much radiated energy wasted or sent to the atmosphere causing radiation noise in the atmosphere. Friis transmission equation is a peculiar equation used in transmission involving communication in terms of antenna towers and fundamentally any form of transmission lines. Optical flow equation is nothing but mathematically an equation of a line plotted in x and y coordinate.

**Results and Conclusion:** We have established implementation of copper annealed and FR4 lossy material of antenna to be used in mobile 5G communication in an area of 30-40 interval of distance. We have successfully been able to formulate our hypotheses and proposed algorithm in establishing communication advantages. The proposed method proves to be quite effective in 5G communication in modern antennas which vary in designs. An efficiency percentage of 60-70 percent was also observed in our method.